\documentclass[3p, 10pt]{elsarticle}
\usepackage{graphicx,amscd,amsmath,amssymb,verbatim}
\usepackage{amsfonts,epsfig}
\usepackage{mathptmx}
\usepackage{setspace}
\usepackage{epsfig}
\usepackage{url}
\usepackage{multirow}

\begin{document}

\journal{Elsevier}

\begin{frontmatter}

\title{Novel considerations about the non-equilibrium regime of the
tricritical point in a metamagnetic model: localization and tricritical
exponents}

\author{Roberto da Silva$^{1}$}
\author{Henrique A. Fernandes$^{2 }$}
\author{J. R. Drugowich de Fel\'{\i}cio$^{3 }$} 
\author{Wagner Figueiredo$^{4}$ }

\address{1 - Instituto de F\'{\i}sica\\
Universidade Federal do Rio Grande do Sul\\
Avenida Bento Gon\c{c}alves 9500 \\ 
Caixa Postal 15051
 91501-970, Porto Alegre RS, Brazil\\
%Tel: +55 51 3308 6474\\
{\normalsize {E-mail:rdasilva@if.ufrgs.br}}}

\address{2 - Coordena\c{c}\~{a}o de F\'{\i}sica,\\
Universidade Federal de Goi\'{a}s\\ 
Campus Jata\'{\i}, BR 364, km 192, 3800\\
75801-615, Jata\'{\i}, Goi\'{a}s, Brazil\\
{\normalsize{E-mail:ha.fernandes@gmail.com}}}

\address{3 - Departamento de F\'{\i}sica e Matem\'{a}tica,\\ 
Faculdade de Filosofia, Ci\^{e}ncias e Letras de Ribeir\~{a}o Preto,\\ 
Universidade de S\~{a}o Paulo,\\ 
Avenida Bandeirantes, 3900 \\
14040-901, Ribeir\~{a}o Preto, S\~{a}o Paulo, Brazil\\
{\normalsize{E-mail:drugo@usp.br}}}

\address{4- Departamento de F\'{\i}sica,\\ 
Universidade Federal de Santa Catarina,\\
Campus Universit\'{a}rio, Trindade,\\
88040-900 - Florianopolis, Santa Catarina, Brazil\\
{\normalsize{E-mail:wagner@fisica.ufsc.br}}}

%\keywords{1 - Time-dependent Monte Carlo Simulations, 2 - Tricritical Point.
%3 - Metamanetic model, 4- Refinement of critical parameters}

\begin{abstract}

We have investigated the time-dependent regime of a two-dimensional
metamagnetic model at its tricritical point via Monte Carlo simulations.
First of all, we obtained the temperature and magnetic field corresponding to the
tricritical point of the model by using a refinement process based on
optimization of the coefficient of determination in the log-log fit of
magnetization decay as function of time. With these estimates in hand, we
obtained the dynamic tricritical exponents $\theta $ and $z$ and the static
tricritical exponents $\nu $ and $\beta $ by using the universal power-law
scaling relations for the staggered magnetization and its moments at early
stage of the dynamic evolution. Our results at tricritical point confirm
that this model belongs to the two-dimensional Blume-Capel model
universality class for both static and dynamic behaviors, and also they
corroborate the conjecture of Janssen and Oerding for the dynamics of
tricritical points.

\end{abstract}

\end{frontmatter}

\setlength{\baselineskip}{0.5cm}

In the study of phase transitions and critical phenomena, systems which
exhibit multicritical behavior have been the subject of a great number of
works. Theoretically, the tricritical phase transition of the Blume-Capel 
\cite{Blume1966} model is one of the most studied. However, there are other
models showing the existence of such multicritical points, for instance, the
metamagnetic model \cite{Motizuki1959}, the Blume-Capel model with
antiferromagnetic exchange interaction and external magnetic field added 
\cite{Wang1976}, and the random-field Ising model \cite{Kaufman1986}. In
order to investigate these phenomena, several techniques have been employed,
including series expansions \cite{Saul1974}, linked-cluster expansion \cite%
{Wang1984}, mean-field theory \cite{Furman1977}, renormalization group \cite%
{Burkhardt1977,Yeomans1981,Bakchich1992,deOliveira1995}, transfer matrix 
\cite{Rikvold1983,Herrmann1984,Alcaraz1985,Beale1986}, Monte Carlo
simulations \cite{Jain1980,Kimel1987,Wang1991,Ayat1993}, and Monte Carlo
renormalization group methods \cite{Landau1981,Landau1986,Honda1993}.
Experimentally, the phase transitions of metamagnetic systems such as in the
compound FeBr$_{2}$ \cite{Katsumata1997,Petracic1997} have also been studied
in order to understand the tricritical behavior that appears as a
consequence of a competition between the antiferromagnetic and ferromagnetic
coupling constants present in this magnetic system.

The two-dimensional spin$-\frac{1}{2}$ metamagnetic model is defined by the
Hamiltonian 
\begin{equation}
\mathcal{H}=J_{1}\sum_{nn}\sigma _{i}\sigma _{j}-J_{2}\sum_{nnn}\sigma
_{i}\sigma _{k}+H\sum_{i}\sigma _{i}
\end{equation}%
where $J_{1}$, $J_{2}>0$ and $\sigma _{i}=\pm 1$ are the spin variables. The
considered model has two sublattices where the first sum extends over all
nearest-neighbor pairs (intersublattice) and second one over all
next-nearest neighbor pairs (intrasublattice), respectively. The parameters $%
J_{1}$ and $J_{2}$ are the antiferromagnetic and ferromagnetic coupling
constants, respectively, and $H$ is the external magnetic field.

The order parameter of the model is the staggered magnetization,
conveniently defined by 
\begin{equation}
M(t)=\frac{1}{N}\sum_{i=1}^{L}\sum_{j=1}^{L}(-1)^{i+j}\sigma
_{i,j}=M_{1}(t)-M_{2}(t),  \label{Eq:OP}
\end{equation}%
where $N=L^{2}$, $L$ is the linear size of the square lattice. Here $%
M_{1}(t)=\frac{2}{N}\sum_{i=1}^{L}\sum_{j=1}^{L}\sigma _{i,j}$ $\delta _{%
\text{mod}(i+j,2),0}$ and $M_{2}(t)=\frac{2}{N}\sum_{i=1}^{L}\sum_{j=1}^{L}%
\sigma _{i,j}$ $\delta _{\text{mod}(i+j,2),1}$ denote the magnetizations of
the respective sublattices. This definition shows that there is an inversion
of the meaning of ordered and disordered state. In order to obtain an
ordered state, it is necessary to occupy the sites of the lattice with spins 
$+1$ $(-1)$ where the sum $i+j$ is odd (even), or vice-versa. On the other
hand, the null magnetization may be obtained when all sites are occupied
with the spins of the same kind.

On the contrary of the Blume-Capel model, the phase diagram of the
metamagnetic model has not yet been completely understood. This is due to
the controversial results between the experimental and theoretical works
concerning the phase transitions of the system. If on the one hand, this
model exhibits a rich phase diagram in the temperature-field plane with a
line of second-order phase transitions, a line of first-order phase
transitions and a tricritical point which is located at the point where the
first and second order transition lines join each other with the same slope,
one the other hand the mean-field theory \cite{Kincaid1975} predicts that
such tricritical point depends on the value of the ratio between the
coupling constants. They only predicted the existence of a tricritical point
for $R=J_{2}/J_{1}>3/5$, while for $R<3/5$ \ in the mean field approximation
the model exhibits two Ising-like critical points: a critical endpoint
corresponding to a point that ends at the first order line coming from the
second order line and a double critical endpoint (bicritical) that
corresponds to the terminal point of the first order transition line.
Although for the three-dimensional metamagnetic model Herrmann et al. \cite%
{Hermann1982} showed via Monte Carlo Renormalization group that such
critical endpoints exist, experimental works have not found those points in
any real metamagnetic system, and also there is no evidence of such points
for the two-dimensional metamagnetic systems as verified in different works
(see for example \cite{Herrmann1984}, \cite{Beale1984}). Similarly, Santos e
Figueiredo \cite{Santos1998} by using master equation formalism on the
context of dynamical pair aproximation, also in two dimensions, did not find
any evidence for the decomposition of the tricritical point into the
critical and bicritical end points as predicted by the mean field theory.
More recently, other authors exclude the possibility of existence of these
two critical endpoints even for three dimensions -- Geng et al. \cite%
{Geng2008}, by using effective field theory, showed that there is no
fourth-order critical point or reentrant phenomenon in the phase diagram.
Finally, other authors \cite{Zukovic2013} by performing MC simulations
showed that there is no evidence of such a decomposition in a critical
endpoint and a bicritical endpoint and such simulations produce a
tricritical behaviour even for a coupling ratio as small as $R=0.01$.\textbf{%
\ }

Although the previous estimates of the critical exponents for this model
support the assertion that it belongs to the same universality class of the
Blume-Capel model, the nonequilibrium critical behavior of this system has
not been completely investigated up to date. Santos and Figueiredo \cite%
{Santos2000} studied a similar layered metamagnetic model far from
equilibrium by using short-time Monte Carlo simulations. They estimated the
static critical exponents $\beta $ and $\nu $ and the dynamic critical
exponent $z$ on the continuous transition line, but the tricritical
exponents were not obtained. They also showed that although the critical
exponent $\nu $ remains the same along the continuous transition line, the
exponent $\beta $ departs from the expected value as we approach the
tricritical point of the model.

The study of the dynamic critical properties of statistical systems has been
a subject of considerable interest in non-equilibrium physics after the
works by Janssen, Schaub and Schmittmann \cite{Janssen1989}, and Huse \cite%
{Huse1989}. By using, respectively, renormalization group techniques and
numerical calculations, they showed that universality and scaling behavior
are already present in systems since their early stages of the time
evolution after quenching from high temperatures to the critical one. As a
result, the study of the critical properties of statistical systems became
in some sense simpler, because they allow to circumvent the well-known
problem of critical slowing down, characteristic of the long-time regime.

The dynamic scaling relation obtained by Janssen \textit{et al.} for the 
\textit{k}-th moment of the magnetization $M$, extended to systems of finite
size \cite{Li1995,Zheng1998}, is written as 
\begin{equation}
\langle M^{k}\rangle (t,\tau ,L,m_{0})=b^{-k\beta /\nu }\langle M^{k}\rangle
(b^{-z}t,b^{1/\nu }\tau ,b^{-1}L,b^{x_{0}}m_{0}),  \label{Eq:Main_short_time}
\end{equation}%
where $t$ is the time, $b$ is an arbitrary spatial rescaling factor, $\tau
=\left( T-T_{c}\right) /T_{c}$ is the reduced temperature and $L$ is the
linear size of the lattice. Here, the operator $\langle \ldots \rangle $
denotes averages over different configurations due to different possible
time evolution from each initial configuration compatible with a given
initial magnetization $m_{0}$. The exponents $\beta $ and $\nu $ are the
equilibrium critical exponents associated to the order parameter and the
correlation length respectively, and $z$ is the dynamic exponent
characterizing time correlations at equilibrium.

After choosing the scaling $b^{-1}L=1$ at the $T=$ $T_{c}$ ($\tau =0$), and $%
k=1$, we obtain $\langle M\rangle (t,L,m_{0})=L^{-\beta /\nu }\langle
M\rangle (L^{-z}t,L^{x_{0}}m_{0})$. Denoting $u=tL^{-z}$ and $%
w=L^{x_{0}}m_{0}$, one has: $\langle M\rangle (u,w)=\langle M\rangle
(L^{-z}t,L^{x_{0}}m_{0})$. The derivative with respect to $L$ is: 
\begin{eqnarray*}
\frac{\partial \langle M\rangle }{\partial L} &=&(-\beta /\nu )L^{-\beta
/\nu -1}\langle M\rangle (u,w)+ \\
&&L^{-\beta /\nu }\left[ \frac{\partial \langle M\rangle }{\partial u}\frac{%
\partial u}{\partial L}+\frac{\partial \langle M\rangle }{\partial w}\frac{%
\partial w}{\partial L}\right] ,
\end{eqnarray*}%
where $\partial u/\partial L=-ztL^{-z-1}$ and $\partial w/\partial
L=x_{0}m_{0}L^{x_{0}-1}$. In the limit $L\rightarrow \infty $, $\partial
_{L}\langle M\rangle \rightarrow 0$, one has: $x_{0}w\frac{\partial \langle
M\rangle }{\partial w}-zu\frac{\partial \langle M\rangle }{\partial u}-\beta
/\nu \langle M\rangle =0$. The separability of the variables $u$ and $w$ in $%
\langle M\rangle (u,w)=M_{1}(u)M_{2}(w)$ leads to $x_{0}wM_{2}^{\prime
}/M_{2}=\beta /\nu +zuM_{1}^{\prime }/M_{2}$, where the prime means the
derivative with respect to the argument. Since the left-hand side of this
equation depends only on $w$ and the right-hand side depends only on $u$,
they must be equal to a constant $c$. Thus, $M_{1}(u)=u^{(c/z)-\beta /(\nu
z)}$ and $M_{2}(w)=w^{c/x_{0}}$, resulting in $\left\langle M\right\rangle
(u,w)=m_{0}^{c/x_{0}}L^{\beta /\nu }t^{(c-\beta /\nu )/z}$. Returning to the
original variables, one has: $\langle M\rangle
(t,L,m_{0})=m_{0}^{c/x_{0}}t^{(c-\beta /\nu )/z}$.

On one hand, choosing $c=x_{0}$ and denoting $\theta =(x_{0}-\beta /\nu )/z$%
, at criticality ($\tau =0$), we obtain the algebraically behavior of the
magnetization, 
\begin{equation}
\langle M\rangle (t)\thicksim m_{0}t^{\theta }.  \label{Eq:theta}
\end{equation}

This can be observed by a finite time scaling $b=t^{1/z}$, Eq. (\ref%
{Eq:Main_short_time}), at critical temperature ($\tau =0$), which leads to $%
\left\langle M\right\rangle (t,m_{0})=t^{-\beta /(\nu z)}\langle M\rangle
(1,t^{x_{0}/z}m_{0})$. Defining $x=t^{x_{0}/z}m_{0}$, an expansion of the
averaged magnetization around $x=0$ results in: $\langle M\rangle
(1,x)=\langle M\rangle (1,0)+\left. \partial _{x}\langle M\rangle
\right\vert _{x=0}x+\mathcal{O}(x^{2})$. By construction $\langle M\rangle
(1,0)=0$, since $x=t^{x_{0}/z}m_{0}\ll 1$ and $\left. \partial _{x}\langle
M\rangle \right\vert _{x=0}$ is a constant. Discarding the quadratic terms
we obtain the expected power law behavior $\langle M\rangle _{m_{0}}\sim
m_{0}t^{\theta }$, which is valid only for a characteristic time scale $%
t<t_{\max }\sim m_{0}^{-z/x_{0}}$.

Then, in this new universal regime, in addition to the familiar set of
critical exponents described above, a new dynamic critical exponent $\theta $
is found. This exponent, independent of the previously known ones,
characterizes the so called \textquotedblleft critical initial
slip\textquotedblright , the anomalous behavior of the magnetization when
the system is quenched to the critical temperature $T_{c}$. In addition, a
new critical exponent $x_{0}$ is introduced to describe the dependence of
the scaling behavior on the initial conditions. This exponent represents the
anomalous dimension of the initial magnetization $m_{0}$ and is related to
the exponent $\theta $ as $x_{0}=\theta z+\beta /\nu $.

On the other hand, the choice $c=0$ corresponds to the case which the system
does not depend on the initial trace and in which $m_{0}=1$ leads to simple
power law: 
\begin{equation}
\langle M\rangle _{m_{0}=1}\sim t^{-\beta /(\nu z)},  \label{decay_ferro}
\end{equation}%
which corresponds to decay of magnetization at long times ($t>t_{\max }$).

Unlike the second-order phase transition, the behavior of a thermodynamic
system is more complex at a tricritical point and the corresponding exponent 
$\theta $ may assume negative values. This assertion was theoretically
deduced by Janssen and Oerding \cite{Janssen1994} and numerically confirmed
by da Silva \textit{et al.} \cite{daSilva2002} through short-time Monte
Carlo simulations at the tricritical point of the Blume-Capel model.
However, as shown by some researchers, negative values of the exponent $%
\theta $ can be also found in systems exhibiting continuous phase
transitions, as for instance, the Baxter-Wu \cite{Arashiro2003,Malakis2005},
multispin \cite{Simoes2001}, and 4-state Potts models \cite{daSilva2004}.

At a tricritical point, the magnetization shows a crossover from the
logarithmic behavior $M(t)\thicksim m_{0}[\ln (t/t_{0})]^{-a}$ at short
times $t\ll m_{0}^{-4}$ to $t^{-1/4}$ power law with logarithmic
corrections, $M(t)\thicksim \lbrack t/\ln (t/t_{0})]^{-1/4}$ in three
dimensions. The above behavior can be stated in the generalized form \cite%
{Janssen1994} 
\begin{equation}
M(t)=m_{0}\left[ \ln \left( \frac{t}{t_{0}}\right) \right] ^{-a}F_{M}(x),
\label{Eq:magtricritical}
\end{equation}%
where 
\begin{equation}
x=\left\{ \left( \frac{t}{\ln (t/t_{0})}\right) ^{\frac{1}{4}}\left[ \ln
\left( \frac{t}{t_{0}}\right) \right] ^{-a}m_{0}\right\} .
\end{equation}%
In Eq. (\ref{Eq:magtricritical}) the function behave as $F_{M}(x)\thicksim 1$
or $F_{M}(x)\thicksim 1/x$ for vanishing and large arguments, respectively.
Below three dimensions it reduces to the scaling form given by Eq. (\ref%
{Eq:theta}), but now the exponent $\theta $ is the exponent related to the
tricritical point of the relaxation process at early times.

In the present work, the simulations were carried out for square lattices
with linear dimension $L=160$ and periodic boundary conditions for all
performed experiments. The estimates for each exponent were obtained from
five independent bins at the tricritical point, each one consisting of $%
N_{run}$ (number of different time series of magnetization or its upper
moments from an initial configuration) runs and $N_{MC}$ Monte Carlo sweeps.
The error bars are fluctuations of the averages obtained from those bins and
the dynamic evolution of the spins is local and updated by the heat-bath
algorithm. Here it is important to mention that finite size scaling effects
are negligencible for the considered size $L=160$. For example, by using the
Eq. (\ref{decay_ferro}) with $N_{run}=10000$ runs and $L=80$, we obtained $%
\beta /(\nu z)=0.03938(5)$ form the time interval $[80,300]$ while for $%
L=140 $ we obtained $\beta /(\nu z)=0.03934(5)$ for the same interval and
number of runs. So, within statistical errors we can not distinguish the
results for $L=80$ and $L=140$. Then, we are very confortable with $L=160$.
Other details about simulations will be supplied according to development of
this manuscript.

In this paper, we performed time-dependent Monte Carlo (MC) simulations to
explore the tricritical behavior of the two-dimensional metamagnetic model.
First of all, we worked on localization of the tricritical point by using a
recent refinement process developed by da Silva et al. \cite{daSilva2012}.
By considering as input the parameter $\alpha =J_{2}/J_{1}=1/2$, the
supposed tricritical temperature $T_{t}$, the resolution $\Delta H$, $%
N_{MC}=150$ and $N_{run}=1000$ (a large number of runs is not required to
these experiments since relaxation from ordered initial lattices are very
stable, differently from evolutions from disordered initial states which
demand a lot of runs), we performed MC simulations starting always from the
ordered state ($m_{0}=1$) in order to estimate the value of the external
magnetic field $H_{t}$ at the tricritical point. With the tricritical set $%
T_{t}$ and $H_{t}$ in hand, we are then able to estimate some dynamic and
static critical exponents of the proposed model.

To reach our goal, the time evolution of the magnetization (Eq. (\ref%
{decay_ferro})) is obtained for each value of the external magnetic field $%
H_{i}$ in a range $\left[ H_{\min },H_{\max }\right] $, where $H_{i}=H_{\min
}+i\cdot \Delta H$, $i=0,...,n$ , and $n=(H_{\max }-H_{\min })/\Delta H$.
Then, the $H_{t}$ is obtained by using the so-called determination
coefficient of the fit

\begin{equation}
r=\frac{\sum\limits_{t=1}^{N_{MC}}(\overline{\ln \langle M\rangle }-a-b\ln
t)^{2}}{\sum\limits_{t=1}^{N_{MC}}(\overline{\ln \left\langle M\right\rangle 
}-\ln \langle M\rangle (t))^{2}}.  \label{determination_coefficient}
\end{equation}

Here, the closer to the unity is the coefficient, the better is the fit and
the estimate of $H_{t}$. This approach is simpler to calculate than other
schemes, such as the goodness of fit, for example.

In Eq. (\ref{determination_coefficient}) $\overline{\ln \langle M\rangle }%
=(1/N_{MC})\sum\nolimits_{t=1}^{N_{MC}}\ln \langle M\rangle (t)$, $\langle
M\rangle (t)=(1/L^{2})\sum_{j=1}^{N_{run}}M_{j}(t)$, with $M_{j}(t)$
denoting the magnetization of $j$-th run of $t$-th MC step, $a$ and $b$ are
the linear coefficient and the slope in the linear fit $\ln \langle M\rangle 
$ versus $\ln t$, respectively. In our experiments we discarded the initial $%
30$ MC steps for better estimate. In previous work (Ref. \cite%
{daSilva2012,daSilva2013}) we performed two successive refinements: one with
larger $\Delta ^{(0)}$ and another more refined (smaller) $\Delta ^{(1)}$.
Here since we have a good initial kick, we performed only one refinement
with $\Delta H=10^{-3}$.

\begin{figure}[h]
\begin{center}
\includegraphics[width=4.0in]{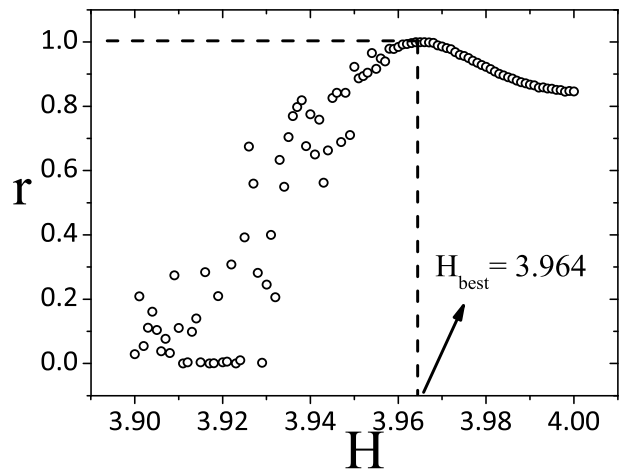}
\end{center}
\caption{Determination coefficient $r$ as function of $H_{t}$ from $H_{\min
}=3.9$ up to $H_{\max }=4.0$ with $\Delta H=0.001$, for the fixed
temperature $T_{t}=1.208$. }
\label{r_versus_field}
\end{figure}

In the first attempt to obtain our set of tricritical parameters ($%
T_{t},H_{t}$), we considered the results obtained by Landau and Swendsen 
\cite{Landau1981}, $T_{t}=1.208(9)$ and $H_{t}=3.965(17)$, which were
obtained through MC renormalization techniques. So, we fixed $T_{t}=1.208$
and changed $H_{t}$ in order to obtain its best value through Eq. (8).
Figure \ref{r_versus_field} shows the plot of $r$ versus $H$ when $H_{\min
}=3.9$ and $H_{\max }=4$.

The best value corresponding to $T_{t}=1.208$ found by our refinement is $%
H_{t}=H_{best}=3.964$ which corroborates the value found in literature.
Since we have observed that our non-equilibrium method is able to determine
the tricritical point, we can now use this estimate to check if, for
instance, the exponent $z$ is consistent with the results of literature. For
this purpose we used a function $F_{2}(t)$ given by \cite{daSilva2002a} 
\begin{equation}
F_{2}(t)=\frac{\langle M^{2}(t)\rangle _{m_{0}=0}}{\langle M(t)\rangle
_{m_{0}=1}^{2}}\thicksim t^{d/{z}},  \label{Eq:f2}
\end{equation}%
where $d$ is the dimension of the system. This approach, that mixes moments
of magnetization under different initial conditions, has proved to be very
efficient in estimating the exponent $z$ for a great number of models \cite%
{daSilva2002,Arashiro2003,Fernandes2005,daSilva2005,daSilva2004a}. Here it
is important to mention that the calculation of the second moment $\langle
M^{2}(t)\rangle _{m_{0}=0}$ must be performed with initial magnetization per
spin $m_{0}=0$. However, the initial configuration of the system must be
choosen at random instead of ordered one. As mentioned above, from Eq. (\ref%
{Eq:OP}), an initial ordered configuration ( $\sigma _{i,j}=1$ for all
sites) gives $M(0)=m_{0}=0$ and, although this is the simpler way, it is a
very correlated one. Nevertheless, the sharp preparation of random initial
condition with $m_{0}=0$ is also straightforward performed: we distribute
randomly, and with same probability, spins $+1$ and $-1$ on the lattice
sites. Then, an adjustment process is performed: If the magnetization is
negative, we choose randomly one site $(i,j)$, and while the magnetization
remains negative, we flip $+\rightarrow -$ whether $i+j$ is odd and $%
-\rightarrow +$ otherwise. If the magnetization is positive we make exactly
\ the opposite: we flip $-\rightarrow +$ if $i+j$ is odd and $+\rightarrow -$
otherwise. This process is done until the magnetization vanishes.

On the other hand, to obtain the magnetization $\left\langle
M(t)\right\rangle $, we must perform simulations with ordered initial
configurations which are trivially preparated by putting in a site $(i,j)$ a
spin $+1$ if $i+j$ is even and $-1$ otherwise.

In this paper, we used for computation of averaged time series of the $k$-th
moment of magnetization, i.e., $\langle M^{k}(t)\rangle _{m_{0}}\times t$, a
total of $N_{run}\ $runs that depends on $m_{0}$ considered. The error bars
were obtained from $N_{b}$ different bins (of course each bin means the
quantities -- magnetization or their moments -- were averaged over the $%
N_{run}$ time series). In order to obtain the tricritical exponents, we used
always $N_{b}=5$, $N_{run}=20000$ for experiments that require disordered
initial configurations, such as those ones to obtain the exponents $\theta $
and $z$ (small or null values of $m_{0}$, respectively), and $N_{run}=10000$
for experiments that demand ordered initial configurations, such as those
ones to estimate the exponents $z$, $\beta $ and $\nu $. Our results in the
plots correspond to more refined estimate $\overline{\langle M^{k}(t)\rangle 
}=(1/N_{b})\sum\nolimits_{i=1}^{N_{b}}\langle M^{k}(t)\rangle ^{(i)}$ and
the error bars (standard deviation of average) were estimated as $\sigma /%
\sqrt{N_{b}}=\left( \frac{1}{N_{b}(N_{b}-1)}\sum\nolimits_{i=1}^{N_{b}}\left[
\langle M^{k}(t)\rangle ^{(i)}-\overline{\langle M^{k}(t)\rangle }\right]
^{2}\right) ^{1/2}$, where $\langle M^{k}(t)\rangle ^{(i)}$ denotes the
average of $k$-th moment of magnetization of the $i$-th bin.

\begin{figure}[h]
\begin{center}
\includegraphics[width=3.5in]{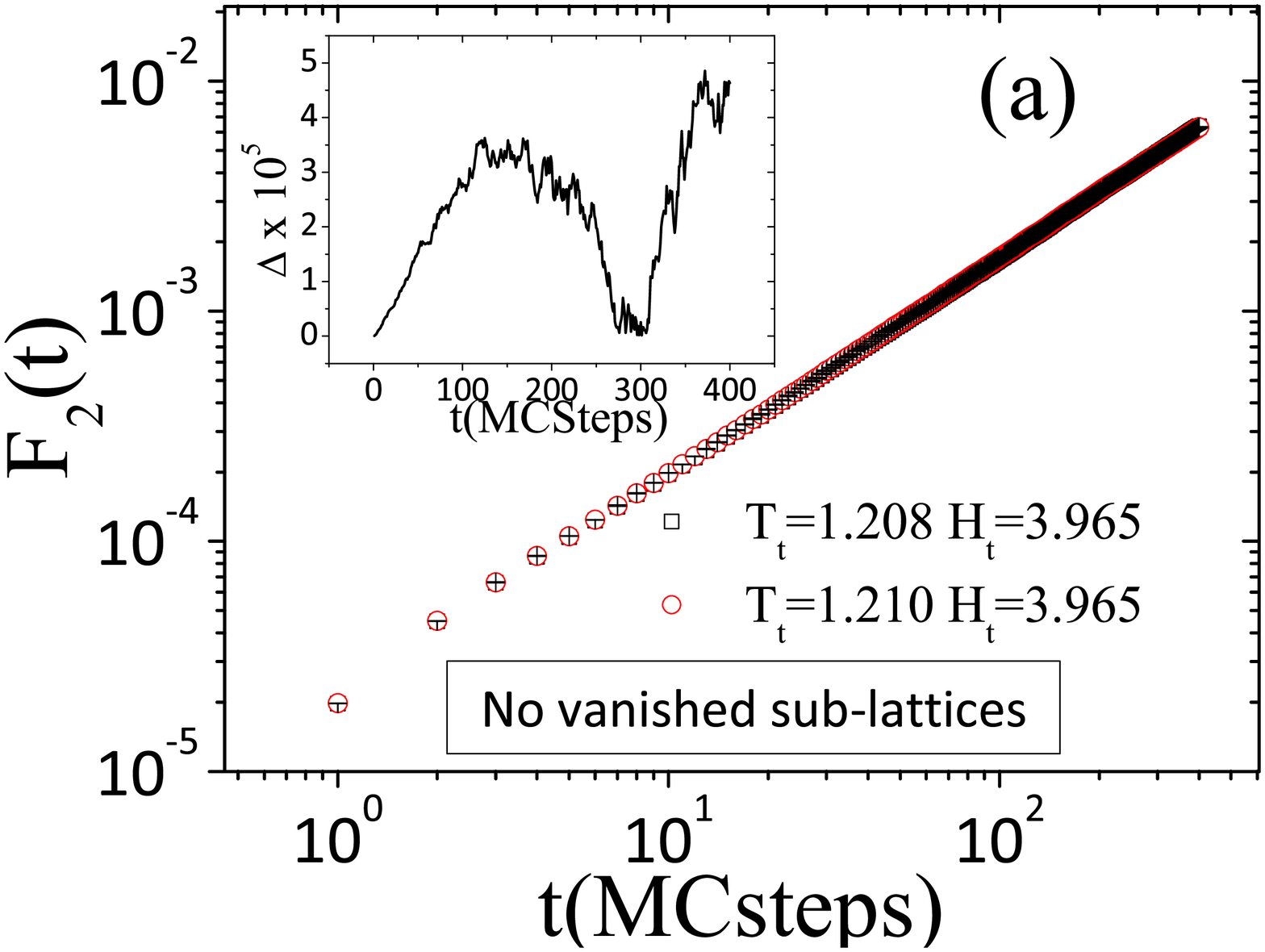}%
\includegraphics[width=3.5in]{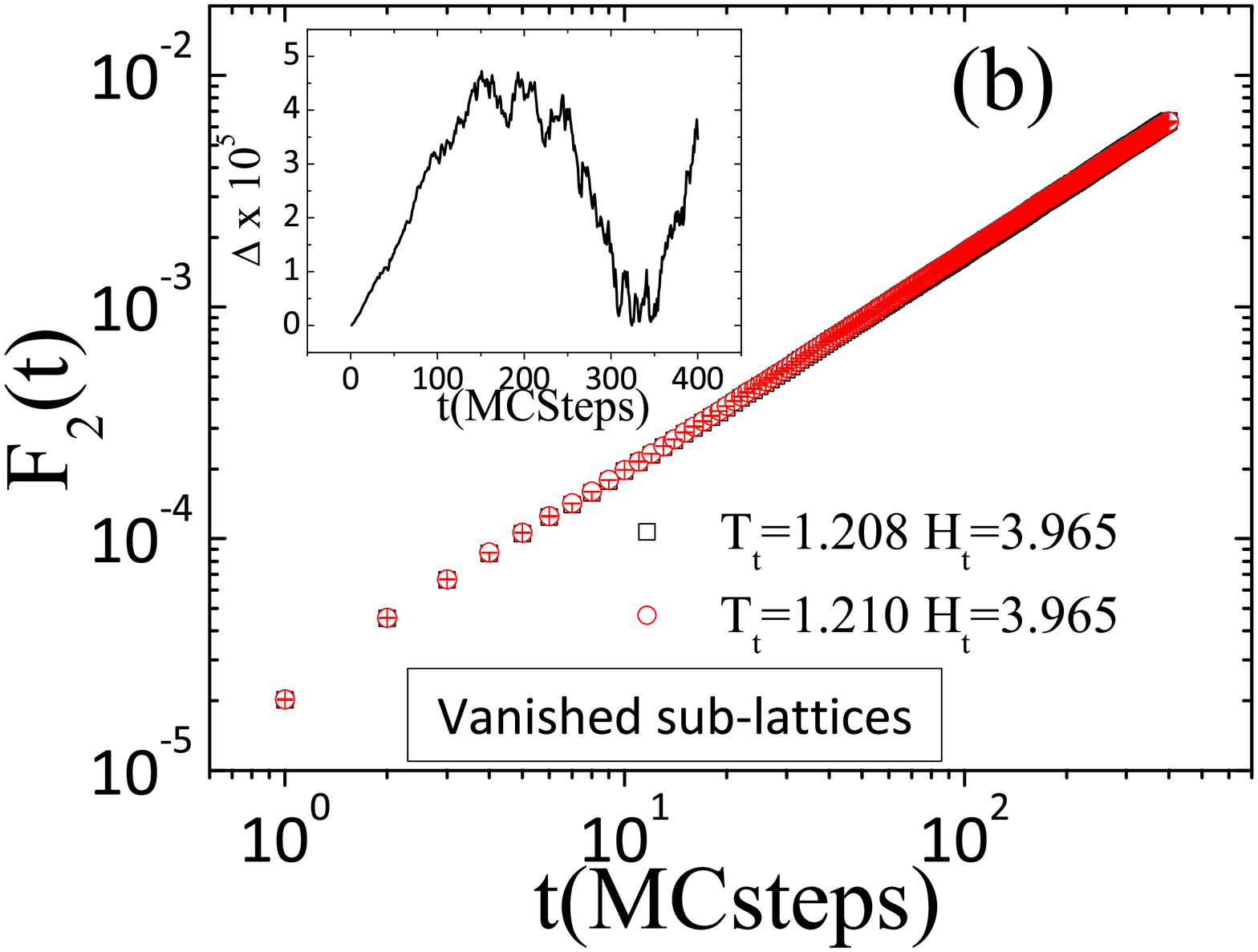}
\end{center}
\caption{\textbf{(a)}: Time evolving in log-log scale of $F_{2}(t)$ as
funtion of time $t$ (corresponding MC step). The square points correspond to
simulations that run set on estimate of Landau and Swendsen ($T_{t}=1.208$
and $H_{t}=3.965$). The circles corresponds to simulations run set on
re-estimated \ with our refinement ($T_{t}=1.210$ and $H_{t}=3.965$). 
\textbf{(b)}: Corresponds to the same plot however the $m_{0}=0$ is obtained
vanishing also the sub-lattices. The same convention is used for both Figure
(a) and (b). The inset plots correspond to behavior of difference in
absolute value between $F_{2}$ obtained with $T_{t}=1.208$ and $F_{2}$
obtained with $T_{t}=1.210$, as function of time. Both cases we used $%
H_{t}=3.965$. }
\label{F2}
\end{figure}

In Figure \ref{F2} (a), we showed the time evolution of $F_{2}(t)$ in a
log-log plot: the black squares correspond to simulations performed with
estimates $T_{t}=1.208$ and $H_{t}=3.965$ obtained in \cite{Landau1981}.
Since $F_{2}(t)$ is obtained from two different time evolutions, we obtain
the exponent $z$ by making a crossover of a bin of $\langle M^{2}(t)\rangle
_{m_{0}=0}$ and another bin of $\langle M(t)\rangle _{m_{0}=1}$, resulting
in $N_{b}=25$ and not simply $N_{b}=5$.

Then we built a simple algorithm that, for different peaces of time
evolution, performs a linear fit and always keeps the same number of points.
We used for such estimates a maximum number of MC steps $N_{MC}=400$. Our
algorithm supplies as output the peace of the time window $[t_{\min
},t_{\max }]$ corresponding to the best goodness of fit \cite{Press1996}, as
well as the value of this goodness ($q$) and corresponding $z$ value
obtained from the slope of $F_{2}$ as function of $t$ in log-log scale,
besides considering the error bars to determination of slope and the needed
error propagation. Here, it is important to mention, that we used the
goodness of fit and not the simple and flexible coefficient of determination 
$(r)$ because the error bars were incorporated to obtain quality of the fit,
as well as the slope and its uncertainty. In this case, $z$ is estimated
with respective uncertainty as $\widehat{z}\pm \sigma _{z}=2/(\widehat{2/z}%
)\pm 2/(\widehat{2/z})^{2}\sigma _{\widehat{2/z}}$, where $\widehat{2/z}$ is
the slope estimated of $F_{2}$ versus $t$ and $\sigma _{\widehat{2/z}}$ the
error obtained of this fit.

In our algorithm, $t_{\min }$ varies from $20$ up to $300$ and $t_{\max }$
from $80$ to $400$, with restriction that $t_{\max }-t_{\min }>60$ MC steps.
As best result, the algorithm supplies $z=2.12(2)$, corresponding to $%
q=0.99999935...$ in $[300,360]$. We fixed for our analysis $20$ points per
interval by adjusting the spacing between the poins. This value is lower
than that estimated for the two-dimensional Blume Capel model at the
tricritical point point, $z=2.215(2)$ \cite{daSilva2002}.

In order to check if we were exactly on the tricritical point, we decided to
reestimate it by using our refinement procedure, as described above. In this
case, we fixed $T_{t}=1.210$, and found $H_{t}=3.965$ (exactly as found in 
\cite{Herrmann1984}) that is a little bit different of that obtained
previously, $H_{t}=3.964$ when $T_{t}=1.208$.

Then, we performed simulations for $F_{2}$ at same conditions but now
considering the new set of tricritical parameters, $T_{t}=1.210$ and $%
H_{t}=3.965$. The red circles in Figure \ref{F2} (a) exhibit such time
behavior. We cannot observe a reasonable difference just by looking at the
plot. Instead we can take the difference between the two estimates of $F_{2}$
(inset Figure \ref{F2} (a)) and observe that there is a "microscopic"
difference. By performing again our algorithm that finds the best interval
of time evolution (corresponding to best goodness of fit) we found $%
z=2.21(2) $, $q=0.9999938...$ and coincidentally for the same time interval $%
[300,360]$. Such result corroborates our estimate for the Blume-Capel model $%
z=2.215(2) $ \cite{daSilva2002}.

%TCIMACRO{\TeXButton{B}{\begin{table*}[tbp] \centering}}%
%BeginExpansion
\begin{table*}[tbp] \centering%
%EndExpansion
\begin{tabular}{lllll}
\hline\hline
Interval & $T_{t}=1.208$ & 
\begin{tabular}{l}
$T_{t}=1.208$ \\ 
zero sublattices \\ 
magnetization%
\end{tabular}
& $T_{t}=1.210$ & 
\begin{tabular}{l}
$T_{t}=1.210$ \\ 
zero sublattices \\ 
magnetization%
\end{tabular}
\\ \hline\hline
\lbrack 220,360] & 
\begin{tabular}{l}
$z=2.109(8)$ \\ 
$q=0.9856$%
\end{tabular}
& 
\begin{tabular}{l}
$z=2.063(6)$ \\ 
$q=0.0885$%
\end{tabular}
& 
\begin{tabular}{l}
$z=2.177(7)$ \\ 
$q=0.9576$%
\end{tabular}
& 
\begin{tabular}{l}
$z=2.124(7)$ \\ 
$q=0.0237$%
\end{tabular}
\\ \hline
\lbrack 240,360] & 
\begin{tabular}{l}
$z=2.105(9)$ \\ 
$q=0.9993$%
\end{tabular}
& 
\begin{tabular}{l}
$z=2.086(7)$ \\ 
$q=0.6588$%
\end{tabular}
& 
\begin{tabular}{l}
$z=2.173(9)$ \\ 
$q=0.9936$%
\end{tabular}
& 
\begin{tabular}{l}
$z=2.156(9)$ \\ 
$q=0.8355$%
\end{tabular}
\\ \hline
\lbrack 280,360] & 
\begin{tabular}{l}
$z=2.12(2)$ \\ 
$q=0.9996$%
\end{tabular}
& 
\begin{tabular}{l}
$z=2.13(2)$ \\ 
$q=0.9998$%
\end{tabular}
& 
\begin{tabular}{l}
$z=2.19(2)$ \\ 
$q=0.9773$%
\end{tabular}
& 
\begin{tabular}{l}
$z=2.19(1)$ \\ 
$q=0.9804$%
\end{tabular}
\\ \hline
\lbrack 300,360] & 
\begin{tabular}{l}
$z=2.12(3)$ \\ 
$q=0.9999$%
\end{tabular}
& 
\begin{tabular}{l}
$z=2.13(2)$ \\ 
$q=0.9999$%
\end{tabular}
& 
\begin{tabular}{|l|}
\hline
$z=2.21(2)$ \\ 
$q=0.9999$ \\ \hline
\end{tabular}
& 
\begin{tabular}{l}
$z=2.17(3)$ \\ 
$q=q=0.9991$%
\end{tabular}
\\ \hline\hline
\end{tabular}
\caption{Estimates of z for different intervals and respective goodness of fit found for each time interval analyzed for
$H_{t}=3.965$. Both situations are analyzed with zero sublattices magnetization (no natural choice) and no 
vanished sub-lattices (natural choice).} \label{Table:interval_z}%
%TCIMACRO{\TeXButton{E}{\end{table*}}}%
%BeginExpansion
\end{table*}%
%EndExpansion

Of course, $m_{0}=$ $0$ does not imply in sub-lattices with zero
magnetization. Although this is an artificial preparation, it is worth to
test this situation by observing the time evolution of $F_{2}$, for
instance. So, we prepared the initial states with the spin variables at each
site chosen at random but with $M_{1}(t=0)=M_{2}(t=0)=m_{0}=0$ in order to
study such effects on the exponent $z$ in comparisom with the straight
preparation.

With the view to obtain such configurations we performed the following
procedure: we randomly selected $L^{2}/4$ spins $\sigma _{i,j}$ with $i+j$
even, and attributed $\sigma _{i,j}=1$. We also randomly selected $L^{2}/4$
spins $\sigma _{i,j}$ with $i+j$ odd, and attributed $\sigma _{i,j}=-1$.
Thereafter, we attribute $\sigma _{i,j}:=-1$ to the remaining spins with $%
i+j $ even and $\sigma _{i,j}:=1$ otherwise. The time evolving under such
conditions for the same parameters that was studied without vanishing the
sub-lattices can be observed in Fig. \ref{F2} (b). Similarly by applying our
algorithm to find the best interval, with its corresponding $z$, we obtained 
$z=2.13(2)$ in $[300,360]$ with $q=0.9998798...$ for $T_{t}=1.208$ and $%
H_{t}=3.965$ and $z=2.17(3)$ in $[320,380]$ with $q=0.9999955...$, for $%
T_{t}=1.210$ and $H_{t}=3.965$.

We conclude that vanishing the sublattices magnetization seems to be not
interesting because the exponent $z$ presents difference from the natural
condition of $m_{0}=0$, even though, for$\ T_{t}=1.210$ and $H_{t}=3.965$,
we find an agreement (according to error bars) of exponents with our best
estimate $z=2.21(2)$, obtained without imposing the vanishing of
sub-lattices, whose value is in absolute agreement with estimate to the
tricritical point found for the Blume Capel. Table \ref{Table:interval_z}
summarizes our main results for $z$ for the different situations. It is also
important to mention that vanishing of the sublattices magnetization, for $%
T_{t}=1.210$ we can find $z=2.21(2)$ for other intervals, for example $%
[280,360]$ and with goodness $q=0.992783...$.

\ Since we have determined the tricritical parameters, as well as the
tricritical exponent $z$, now we can calculate the other tricritical
exponents for the metamagnet model, the dynamic critical exponent $\theta $
and static critical exponents $\beta $ and $\nu $. First, we analyzed the
exponent $\theta $, that here is calculated by two different methods: i) the
straight application of the power law behavior given by Eq. (\ref{Eq:theta})
and ii) by means of the time correlation of the order parameter \cite%
{Tania1998}.

In the first method, the exponent $\theta $ is obtained as a function of the
initial magnetization $m_{0}$. In this case, it is necessary working with a
precise and small value of the initial magnetization in order to obtain $%
\theta (m_{0})$. The asymptotic value of $\theta $ is obtained by
extrapolating the estimates of $\theta $ for various values of $m_{0}$
toward the limit $m_{0}\rightarrow 0$. Our simulations were performed for
four different values of $m_{0}$, $m_{0}=0.02,0.04,0.06$, and $0.08$. Here
we used $L=160$ and the initial configurations were prepared with fixed $%
m_{0}$ and spins randomly selected following the procedures previously
described for $m_{0}=0$ without vanishing the sublattices. The only
difference here is that instead of performing an adjustment to find $m_{0}=0$
we perform the adjustment to obtain the fixed desired magnetization.

\begin{figure}[h]
\begin{center}
\includegraphics[width=4.0in]{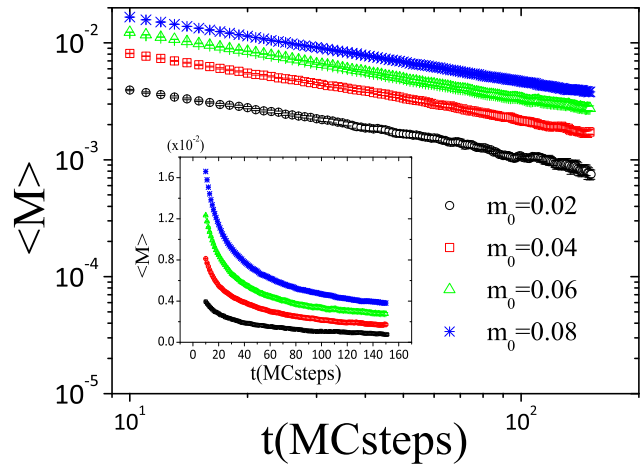}
\end{center}
\caption{Time evolution of the magnetization for $m_{0}=0.02$, $0.04$, $0.06$
and $0.08$. The error bars were calculated over 5 sets of 20000 runs each
one. The inset displays the time evolution in a linear scale.}
\label{Fig:decay_magnetization_m0_small}
\end{figure}

In Figure \ref{Fig:decay_magnetization_m0_small} we showed the behavior of
the time evolution of the staggered magnetization for the considered initial
magnetizations in double-log scale.

\begin{figure}[h]
\begin{center}
\includegraphics[width=4.0in]{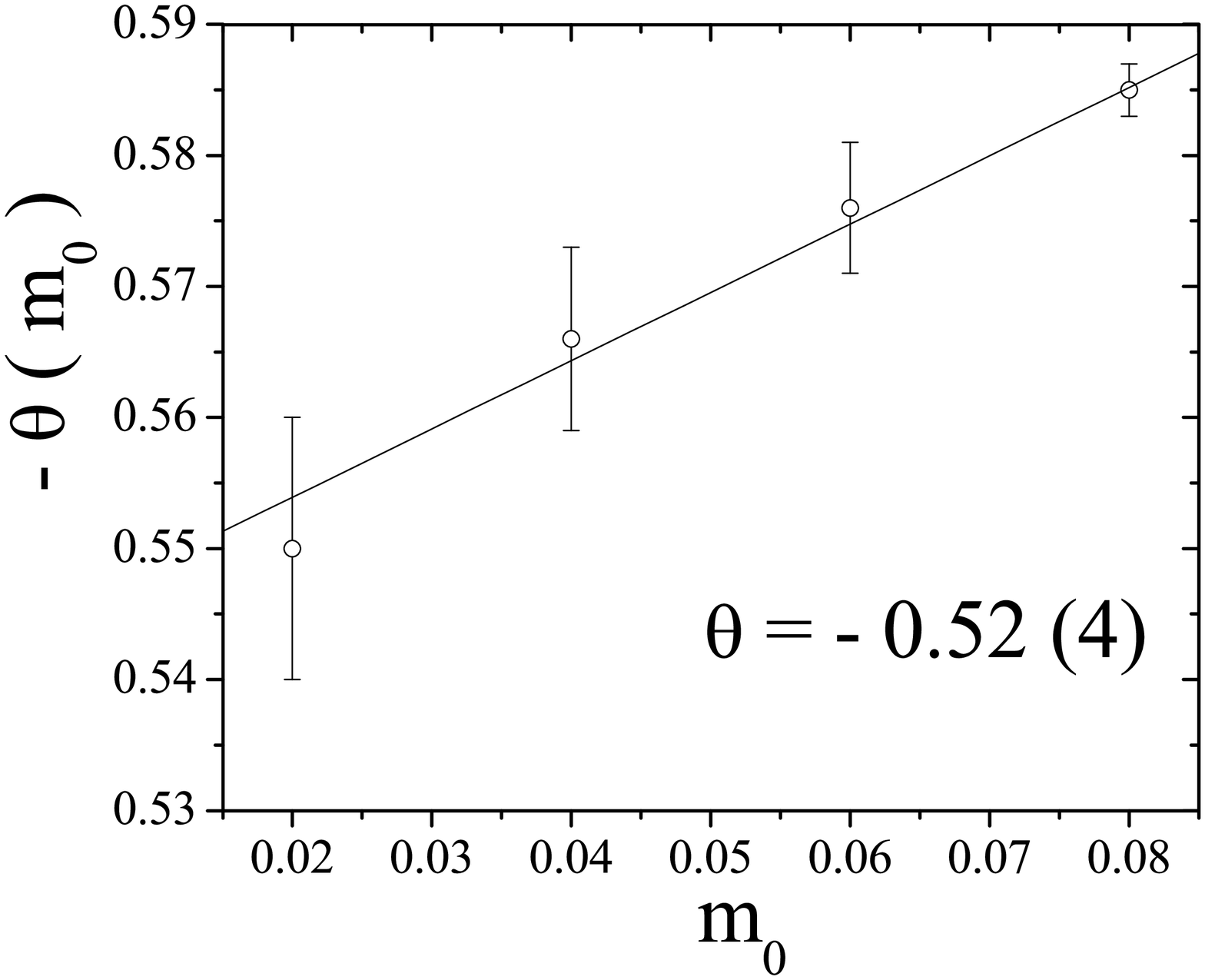}
\end{center}
\caption{Dynamic exponent $\protect\theta $ as a function of the initial
magnetization $m_{0}$. Each point represents an average over 5 sets of 20000
runs each one.}
\label{Fig:Extrapolation}
\end{figure}

Figure \ref{Fig:Extrapolation} exhibits the behavior of the exponent $\theta 
$ for the four initial magnetizations described above, as well as a linear
fit that leads to its final value through the numerical extrapolation
towards $m_{0}\rightarrow 0$. The anomalous behavior which prescribes that
the magnetization decays as a function of time, instead of an expected
increase (as observed in regular critical points) corroborates the numerical
observation of the tricritical point of the Blume Capel model \cite%
{daSilva2002} as well as the theoretical one \cite{Janssen1994}.

In Table \ref{Tb:limit2} we present the estimates for $\theta $ as a
function of different initial magnetizations $m_{0}$. The value found $%
\theta =-0.52(4)$, from extrapolation to $m_{0}\rightarrow 0$.

%TCIMACRO{\TeXButton{B}{\begin{table}[tbp] \centering}}%
%BeginExpansion
\begin{table}[tbp] \centering%
%EndExpansion
\begin{tabular}{ccc}
\hline\hline
$m_{0}$ &  & ~ $\theta $ ~~ \\ \hline\hline
$0.08$ &  & $-0.585(2)$ \\ 
$0.06$ &  & $-0.576(5)$ \\ 
$0.04$ &  & $-0.566(1)$ \\ 
$0.02$ &  & $-0.55(1)$ \\ 
Extrapolated value &  & $-0.52(4)$ ~ \\ \hline\hline
\end{tabular}%
\caption{The dynamical exponent $\theta $ from the time evolution of the
magnetization for different initial configurations.}\label{Tb:limit2}%
%TCIMACRO{\TeXButton{E}{\end{table}}}%
%BeginExpansion
\end{table}%
%EndExpansion

The second method used to estimate $\theta $ is through the time correlation
of the magnetization \cite{Tania1998} given by

\begin{equation}
C(t)=\langle M(0)M(t)\rangle \thicksim t^{\theta }\text{.}
\label{Eq:Correlation}
\end{equation}%
When compared to the first technique foreseen by Eq. (\ref{Eq:theta}), this
method has at least two advantages. It does not demand a careful preparation
of the initial configurations the limiting procedure, the only requirement
being that $\langle m_{0}\rangle =0$. From a computational point of view it
is very useful, because the spins are placed randomly on the lattice sites
and the evolution starts without questions about the value of $m_{0}$ since
that procedure ensures magnetizations around $m_{0}=0$.

\begin{figure}[h]
\begin{center}
\includegraphics[width=4.0in]{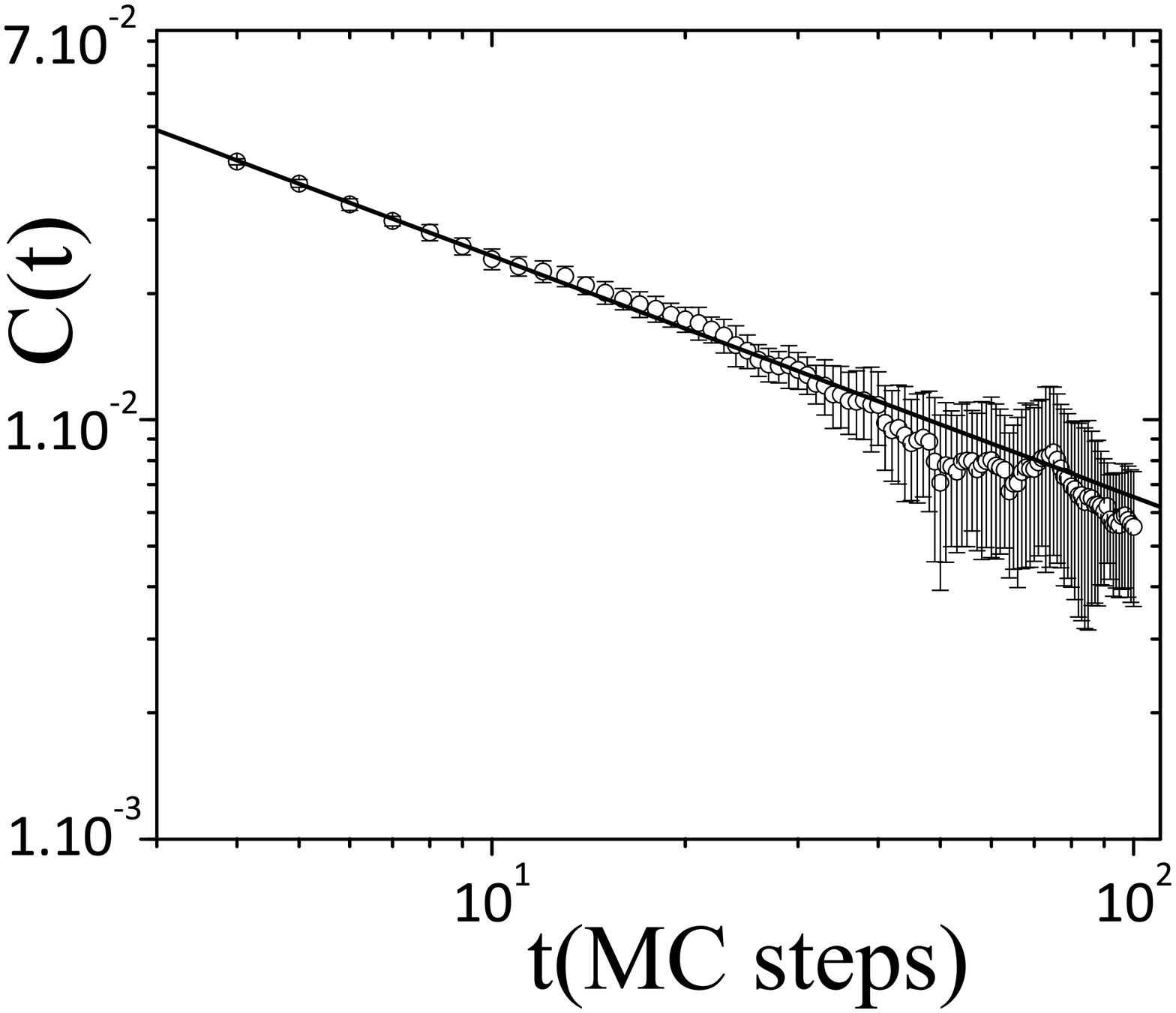}
\end{center}
\caption{Time correlation of the magnetization for samples with $\langle
M(t=0)\rangle \approx 0$. The error bars were calculated over 5 sets of
20000 runs each one.}
\label{Fig:correlation_tome}
\end{figure}

Figure \ref{Fig:correlation_tome} displays the time dependence of the time
correlation $C(t)$ in double-log scale. The linear fit of this curve leads
to the value

\begin{equation}
\theta =-0.56(2).  \label{Eq:restheta}
\end{equation}

These results [see Table \ref{Tb:limit2} and Eq. (\ref{Eq:restheta})] are in
agreement with the value obtained for the Blume-Capel model \cite%
{daSilva2002} at the tricritical point, $\theta =-0.53(2)$, corroborating
the dynamical universality for the tricritical points, as well as confirming
the conjecture by Janssen and Oerding \cite{Janssen1994} whereas the value
of $\theta $ for the metamagnetic model is also negative.

Let us consider now the static critical exponents $\nu $ and $\beta $ of the
metamagnetic model, both obtained through the scaling behavior of the
staggered magnetization and taking into account runs with ordered initial
configurations $(m_{0}=1)$. The statical exponent $\nu $ can be obtained by
fixing $b^{-z}t=1$ in Eq. (\ref{Eq:Main_short_time}) and differentiating $%
\ln M(t,\tau )$ with respect to $\tau $ at the tricritical point. The power
law obtained is 
\begin{equation}
D(t)=\frac{\partial }{\partial \tau }\ln \left\langle M\right\rangle
_{m_{0}=1}(t,\tau )|_{\tau =0}\thicksim t^{1/\nu z}.  \label{Eq:dm}
\end{equation}

\begin{figure}[h]
\begin{center}
\includegraphics[width=4.0in]{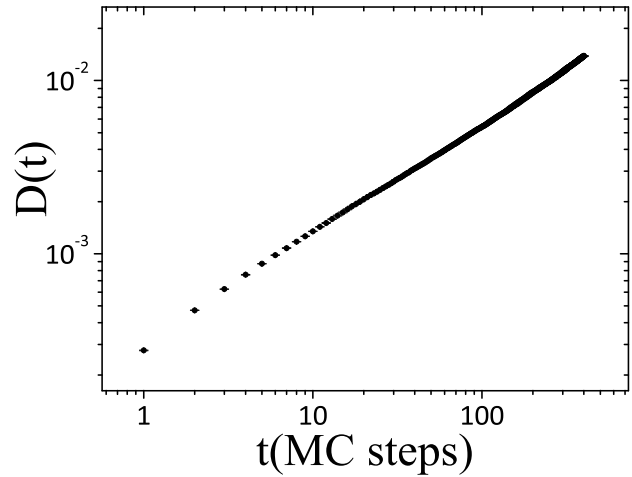}
\end{center}
\caption{The time evolution of the derivative $\partial M(t,\protect\tau %
)/\partial \protect\tau |_{\protect\tau =0}$ in log-log scale in a dynamic
process starting from an ordered state $(m_{0}=1)$. Error bars are smaller
than the symbols. Each point represents an average over 25 sets (5 at $T_{c}-%
\protect\varepsilon $ crossed with 5 at $T_{c}+\protect\varepsilon $) of
10000 runs each one.}
\label{Fig:dvd2}
\end{figure}

Numerically, the quantity $D(t)$ is computed simulating the relaxation of
the system initially ordered in two different points, the first one slightly
above the tricritical temperature $(T_{t}+\varepsilon $) and the other one
slightly below the tricritical temperature $(T_{t}-\varepsilon $), keeping
the magnetic field $H_{t}$ fixed. So, we numerically expected that 
\begin{equation*}
D(t)=\frac{1}{2\varepsilon }\ln \frac{\left\langle M\right\rangle
_{m_{0}=1}(t,\tau +\varepsilon )}{\left\langle M\right\rangle
_{m_{0}=1}(t,\tau -\varepsilon )}\thicksim t^{1/\nu z}
\end{equation*}%
for small value of $\varepsilon $. In previous works we tipically used $%
\varepsilon =O(10^{-3})$ and in this work we considered $\varepsilon =1\cdot
10^{-3}$. In Figure \ref{Fig:dvd2}, the power law increase of the above
equation is plotted in double-log scale.

From the slope of the curve one can estimate the tricritical exponent $1/\nu
z$ and by using the exponent $z$ obtained from the scaling relation $%
F_{2}(t) $, the exponent $\nu $ is then estimated with its respective
uncertainty (error propagated):%
\begin{equation*}
\begin{array}{lll}
\widehat{\nu }\pm \sigma _{\nu } & = & \left( \widehat{1/\nu z}\cdot 
\widehat{z}\right) ^{-1}\pm \\ 
&  & \sqrt{\left( \widehat{1/\nu z}^{2}\cdot \widehat{z}\right) ^{-2}\sigma
_{\widehat{1/\nu z}}^{2}+\left( \widehat{1/\nu z}\cdot \widehat{z}%
^{2}\right) ^{-2}\sigma _{\widehat{z}}^{2}}.%
\end{array}%
\end{equation*}

Here $\widehat{1/\nu z}$ and $\widehat{z}$ are the estimates and $\sigma _{%
\widehat{1/\nu z}}$ and $\sigma _{\widehat{z}}$ are their respective
uncertainties. Our estimate for $\nu $ at the tricritical point is $\nu
_{t}=0.57(3)$ at interval [$320,380$] with goodness-of-fit $q=1$ which
corroborates the theoretical prediction $\nu _{t}=5/9=0.55\overline{5}$.
Here we kept 20 points per interval and used the same processing to find the
best goodness. Here it is also important to mention that once we have
simulated 5 different bins for $T_{t}-\varepsilon $ $(T=1.209)$ and 5 bins
for $T_{t}+\varepsilon \ (T=1.211)$ and by crossing all seeds, we obtained a
sample with 25 different measures as well as the procedure used for $F_{2}$,
i.e., in the first case we have crossed seeds of different temperatures and
in the second case due to the different initial conditions to compose $F_{2}$%
.

Finally, we evaluated the statical exponent $\beta $ by the dynamic scaling
law for the magnetization $\left\langle M\right\rangle
_{m_{0}=1}(t)\thicksim t^{-\beta /{\nu z}}$. By estimating $\widehat{\beta /{%
\nu z}}$ from the log-log plot of $\left\langle M\right\rangle $ versus $t$
, $\beta $ is determined with its uncertainty as

\begin{equation*}
\begin{array}{l}
\widehat{\beta }\pm \sigma _{\beta }=\left( \widehat{\beta /\nu z}\right)
\cdot \left( \widehat{1/\nu z}\right) ^{-1}\pm \\ 
\sqrt{\left( \widehat{1/\nu z}\right) ^{-2}\sigma _{\widehat{\beta /\nu z}%
}^{2}+\left( \left( \widehat{\beta /\nu z}\right) \cdot \left( \widehat{%
1/\nu z}\right) ^{-2}\right) ^{2}\sigma _{\widehat{1/\nu z}}^{2}}%
\end{array}%
\end{equation*}

\begin{figure}[h]
\begin{center}
\includegraphics[width=4.0in]{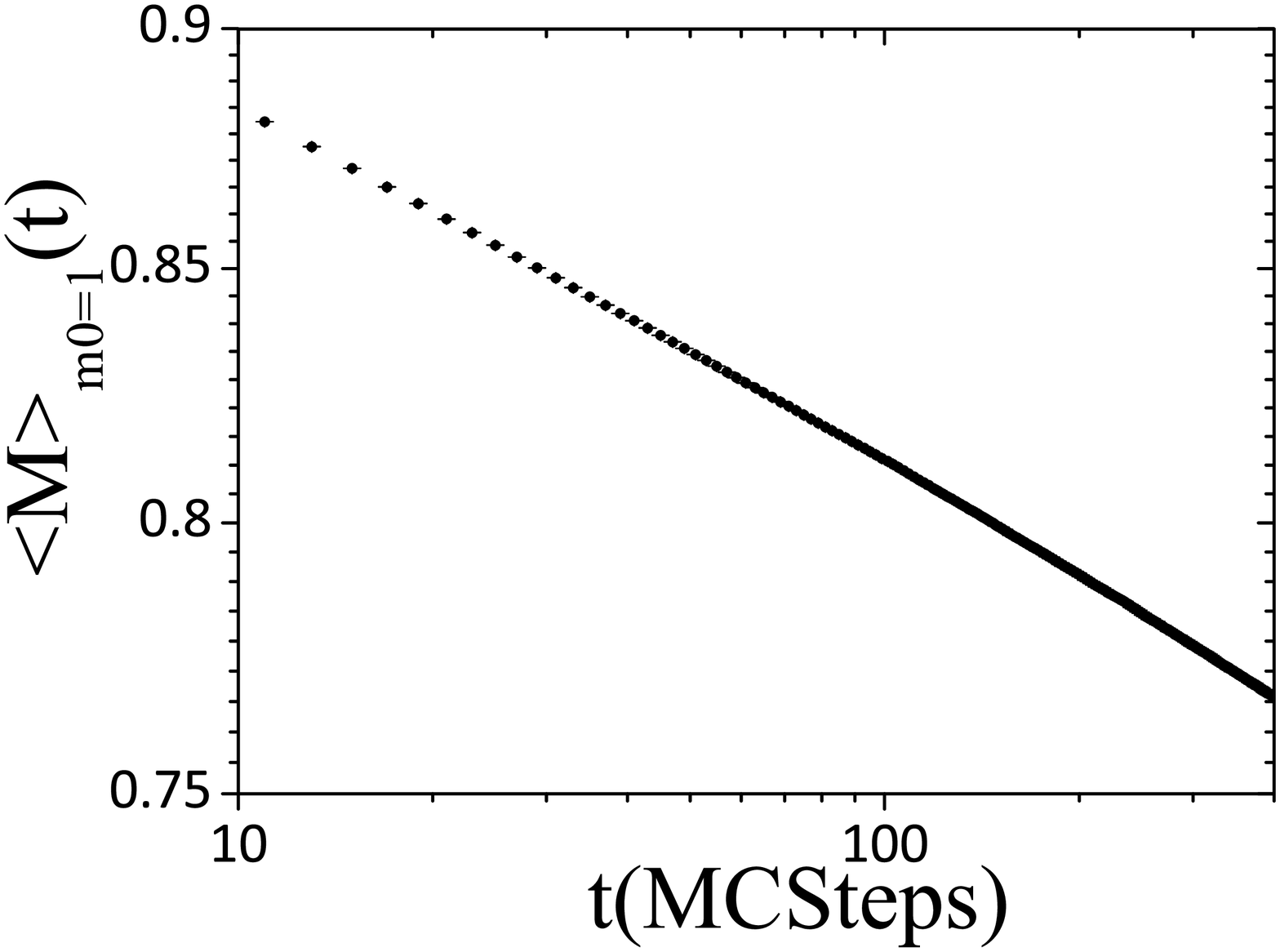}
\end{center}
\caption{The time evolution of the magnetization for initially ordered
samples $(m_{0}=1)$. The error bars calculated over 5 sets of 10000 runs
each one, are smaller than the symbols.}
\label{fig:magord}
\end{figure}

In Figure \ref{fig:magord} we show the time evolution of the magnetization
in double-log scale.

The exponent obtained from the slope of this curve is $\beta /\nu z=0.0390(2)
$ in the time interval $[320,380]$ with $q=0.9999$.... With this exponent in
hand and taking into account the previous result for $1/\nu z\ $($0.79(7)$),
we can estimate the exponent $\beta $ through the equation above. The
result, $\beta =0.049(4)$, is close to the theoretical prediction $\beta
=1/24$. In Table \ref{Table:final} we show our estimates and a comparison
with important results from literature of statical exponents ($\beta $ and $%
\nu $) at the tricritical points.

%TCIMACRO{\TeXButton{B}{\begin{table*}[tbp] \centering}}%
%BeginExpansion
\begin{table*}[tbp] \centering%
%EndExpansion
\begin{tabular}{c|ccccccc}
\hline\hline
Exponent & 
\begin{tabular}{l}
ref. \cite{daSilva2002} \\ 
2D Blume \\ 
Capel \\ 
(short time \\ 
dynamics)%
\end{tabular}
& 
\begin{tabular}{l}
ref. \cite{Alcaraz1985} \\ 
Quantum 1d \\ 
BEG \\ 
Model \\ 
(FSS)%
\end{tabular}
& 
\begin{tabular}{l}
ref. \cite{Landau1981} \\ 
Blume \\ 
Capel \\ 
model \\ 
(MCRG)%
\end{tabular}
& 
\begin{tabular}{l}
ref. \cite{Rikvold1983} \\ 
(FSS)%
\end{tabular}
& 
\begin{tabular}{l}
ref. \cite{Herrmann1984} \\ 
(FSS)%
\end{tabular}
& 
\begin{tabular}{l}
ref. \cite{Balbao1987,Friedan1984} \\ 
(CI)%
\end{tabular}
& 
\begin{tabular}{l}
This \\ 
work%
\end{tabular}
\\ \hline\hline
$\beta $ & $0.0453(2)$ & -- & $0.039$ & $0.0411(7)$ & -- & $1/24$ & $0.049(4)
$ \\ 
&  &  &  &  &  &  &  \\ 
$\nu $ & $0.537(6)$ & $1/1.80$ & $0.56$ & $0.552(6)$ & $0.5562(12)^{\ast }$
& $5/9$ & $0.57(3)$ \\ \hline\hline
\end{tabular}
\caption{Static tricritical exponents. We present some results found in the literature and our predictions 
via time-dependent MC simulations.  The acronyms (BEG), (MCRG), (FSS) and (CI) mean 
"Blume-Emery-Griffiths", "Monte Carlo  "Renormalization Group", Finite Size Scalling, 
and Conformal Invariance respectively.} \label{Table:final}%
%TCIMACRO{\TeXButton{E}{\end{table*}}}%
%BeginExpansion
\end{table*}%
%EndExpansion

In summary, we have performed short-time Monte Carlo simulations to
investigate the scaling behavior at the tricritical point of a
two-dimensional metamagnetic model. The dynamic critical exponent $\theta $
was estimated using two different approaches: by following the time
evolution of the staggered magnetization and measuring the evolution of the
time correlation function $C(t)$ of the staggered magnetization. On the
other hand, the dynamic critical exponent $z$ was found through the function 
$F_{2}(t)$ which combines simulations performed with different initial
conditions. The static critical exponents $\beta $ and $\nu $ were obtained
through the scaling relations for the staggered magnetization and its
derivative with respect to the temperature at $T_{c}$. Our results are in
good agreement with the exponents previously determined for the tricritical
point of the two-dimensional Blume-Capel model.

\section*{Acknowledments}

The authors are partly supported by the Brazilian Research Council CNPq. R.
da Silva thanks to Prof. Leonardo G. Brunet (IF-UFRGS) for the available
computational resources and support of Clustered Computing (ada.if.ufrgs.br).

\end{document}